\documentclass[conference]{IEEEtran}
\IEEEoverridecommandlockouts
\usepackage{cite}
\usepackage[printonlyused,nolist]{acronym}
\usepackage{amsmath,amssymb,amsfonts}
\usepackage{algorithmic}
\usepackage{graphicx}
\usepackage{textcomp}
\usepackage{xcolor}
\usepackage{booktabs}
\usepackage{multirow}
\usepackage{array}
\usepackage{tikz}
\usepackage{pgfplots}
\usepackage{tikzscale}
\usepackage{subcaption}
\usepackage{gensymb}
\def\BibTeX{{\rm B\kern-.05em{\sc i\kern-.025em b}\kern-.08em
    T\kern-.1667em\lower.7ex\hbox{E}\kern-.125emX}}

\begin{acronym}
    \acro{ANN}{artificial neural network}
    \acro{COSPA}{Complex-valued Spatial Autoencoder}
	\acro{DNN}{Deep Neural Network}
    \acro{DoA}{direction of arrival}
    \acro{DSB}{delay-and-sum beamformer}
    \acro{EaBNet}{Embedding and Beamforming Network}
    \acro{ESTOI}{extended Short-Time Objective Intelligibility}
	\acro{GRU}{gated recurrent unit}
    \acro{MC}{multichannel}
    \acro{MVDR}{minimum variance distortionless response}
	\acro{PESQ}{Perceptual Evaluation of Speech Quality}
	\acro{ReLU}{rectified linear unit}
	\acro{SDR}{signal-to-distortion ratio}
	\acro{SIR}{signal-to-interference ratio}
	\acro{SNR}{signal-to-noise ratio}
    \acro{STOI}{Short-Time Objective Intelligibility}
	\acro{STFT}{short-time Fourier transform}
\end{acronym}

\begin{document}

\title{Localizing Spatial Information in Neural Spatiospectral Filters%
}

\author{\IEEEauthorblockN{Annika Briegleb \qquad Thomas Haubner \qquad Vasileios Belagiannis \qquad Walter Kellermann\thanks{This work has been accepted to EUSIPCO 2023.}}
\IEEEauthorblockA{\textit{Multimedia Communications and Signal Processing}\\
\textit{Friedrich-Alexander-Universit\"at Erlangen-N\"urnberg, Erlangen, Germany}\\
\{annika.briegleb, thomas.haubner, vasileios.belagiannis, walter.kellermann\}@fau.de}
}

\maketitle

\begin{abstract}
Beamforming for multichannel speech enhancement relies on the estimation of spatial characteristics of the acoustic scene. In its simplest form, the delay-and-sum beamformer (DSB) introduces a time delay to all channels to align the desired signal components for constructive superposition. Recent investigations of neural spatiospectral filtering revealed that these filters can be characterized by a beampattern similar to one of traditional beamformers, which shows that artificial neural networks can learn and explicitly represent spatial structure. Using the \ac{COSPA} as an exemplary neural spatiospectral filter for multichannel speech enhancement, we investigate where and how such networks represent spatial information. We show via clustering that for \ac{COSPA} the spatial information is represented by the features generated by a gated recurrent unit (GRU) layer that has access to all channels simultaneously and that these features are not source- but only direction of arrival-dependent. 
\end{abstract}


\acresetall

\section{Introduction}
\label{sec:intro}
Spatial filtering for speech and audio signals usually relies on signal-independent and -dependent beamformers such as the \ac{DSB} or the \ac{MVDR} beamformer. Currently, (deep) neural networks that can support \cite{Zhang2017, Wang2018, Martin2020, Masuyama2020, Wang2020} or substitute \cite{Halimeh2022, Li2022, Meng2017, Xiao2016, Li2016} conventional beamformers are moving into the focus for speech enhancement and target speech extraction in noisy and reverberant scenarios. We denote the latter approaches as \textit{neural spatiospectral filters}.

The simplest method for spatial filtering in speech enhancement, the \ac{DSB}, solely relies on phase alignment of the signals captured by the different channels to enhance the desired signal. For this, traditional beamformers require knowledge of the \ac{DoA} of the desired source. On the other  hand, for a single desired speaker in the presence of non-speech or nondirectional background noise, neural spatiospectral filters, often do not require this information \cite{Halimeh2022, Li2022}.
While beampatterns demonstrate that neural spatiospectral filters can represent spatial information  \cite{Halimeh2022, Tesch2022}, it still needs to be clarified how the spatial information contained in the multichannel input is processed and represented by these neural spatiospectral filters. 
In this contribution, we investigate where and how spatial information is captured in an exemplary neural spatiospectral filter. For this, we analyze the \ac{COSPA} \cite{Halimeh2022}, which estimates a complex-valued multichannel mask for speech enhancement. While each neural spatiospectral filter has its own architecture design, most filters include temporal processing. We choose \ac{COSPA} because the joint temporal processing of all channels is localized in one \ac{GRU} layer in the bottleneck of the network and the specific architecture of \ac{COSPA} (cf.~Sec.~\ref{sec:method}) allows to analyze the spatial processing conveniently.
Consequently, we investigate the spatial information contained in the features of the network before and after this \ac{GRU} layer as it is the only network part with memory that can modify all channels individually, which is required to align the phases of the different channel signals.

In Sec.~\ref{sec:method}, we discuss the signal model, the architecture of \ac{COSPA}, and why its \ac{GRU} layer is of special interest for spatial filtering. In Sec.~\ref{sec:representation_spatial_info}, we explain how we measure the spatial information contained in the network's features. We describe our experimental setup in Sec.~\ref{subsec:setup} and present and discuss our results in Sec.~\ref{subsec:results}. Sec.~\ref{sec:conclusion} concludes the paper.

\section{Spatial filtering with COSPA}
\label{sec:method}
In this contribution, we consider the task of extracting the signal of a point-like speech source from a multichannel signal recorded by $M$ microphones in a reverberant room using a neural spatiospectral filter that estimates a complex-valued mask for speech enhancement. The filter can thus modify both the magnitude and the phase of the input signal. In the \ac{STFT} domain, we consider the frame-wise single-channel speech signal estimate $\hat{S}(\tau)$, the multichannel signal $X_m(\tau), m=1,...,M$, and the complex-valued multichannel filter mask  $\mathcal{M}_m(\tau)$ estimated by the neural filter. Frames are half-overlapping and indexed by $\tau$ in both the time and the \ac{STFT} domain. We omit the frequency index for brevity. The estimated speech signal is obtained by 
\begin{equation}
    \hat{S}(\tau) = \sum_{m=1}^{M}\mathcal{M}_m(\tau) \cdot X_m(\tau).
\end{equation}

\begin{figure}[t] 
\vspace{2mm}
	\centering
    \input{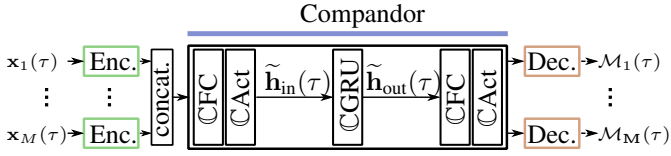}
    \caption{Architecture of COSPA (adapted from \cite{Halimeh2022}).}
    \label{fig:architecture}
\end{figure}

As neural spatiospectral filter for our investigations we choose \ac{COSPA} as introduced in \cite{Halimeh2022}, which estimates the complex-valued mask $\mathcal{M}_m$ for each channel $m$. A simplified structure of \ac{COSPA} is depicted in Fig.~\ref{fig:architecture}. \ac{COSPA} consists of an encoder and a decoder, which both process all signal channels equally, and a compandor, which can process each channel individually. As spatial information is obtained from the differences between signals captured by spatially distributed sensors, and processing it usually includes phase alignment, i.e., delays, the compandor is of specific interest for investigating the representation of spatial information inside \ac{COSPA}.  
As shown in Fig.~\ref{fig:architecture}, the compandor consists of a complex-valued \ac{GRU} layer \cite{Cho2014, Halimeh2021} in between two complex-valued fully connected layers. The \ac{GRU} layer consisting of $U_\text{out}$ complex-valued \acp{GRU} (see \cite{Halimeh2022}) provides memory to the network, which is expected to facilitate the phase alignment of the channels and is thus in the focus of our analysis. The relevance of the \ac{GRU} layer is also supported by \cite{Tesch2022_2}, where the \ac{GRU} layer is initialized with features derived from the \ac{DoA} of the desired speaker to guide the network for target speaker extraction. Note that, here, for the considered scenario of speech enhancement without speech interferers \ac{COSPA} does not receive any explicit information about the \ac{DoA} of the desired source.

As shown in Fig.~\ref{fig:architecture}, we denote the features at the input of the \ac{GRU} layer as $\widetilde{\mathbf{h}}_\text{in}(\tau) = [\widetilde{h}_\text{in}(\tau, 1), ...,\widetilde{h}_\text{in}(\tau, U_\text{in})]^\top$, with $U_\text{in}$ being the number of input channels of the \ac{GRU} layer, and $\widetilde{h}_\text{in}(\tau, u) \in \mathbb{C}$ being a single feature for unit $u$ and time frame $\tau$. These features are obtained from the first fully connected layer in the compandor, which is the first layer in \ac{COSPA} that can fuse all channels. 
Similarly, we define $\widetilde{\mathbf{h}}_\text{out}(\tau) = [\widetilde{h}_\text{out}(\tau, 1), ...,\widetilde{h}_\text{out}(\tau, U_\text{out})]^\top$ as features at the output of the \ac{GRU} layer with $U_\text{out}$ being the number of \acp{GRU} in the \ac{GRU} layer.

\section{Quantifying spatial information}
\label{sec:representation_spatial_info}
As explained in Sec.~\ref{sec:method}, we expect that the features generated by \ac{COSPA}'s \ac{GRU} layer contain spatial information about the desired source. Hence, for a test signal in which the speaker position changes discretely with time, we expect that output features of the \ac{GRU} layer corresponding to frames coming from the same source position are similar to each other and thereby express DoA information resulting from phase alignment. In contrast, at the input of the \ac{GRU} layer we do not expect to find such similarities in the feature vectors. To identify frames characterized by similar feature vectors and to quantify the amount of spatial information contained in the features, we use the $k$-means clustering algorithm \cite{Lloyd1982} on the features $\widetilde{\mathbf{h}}_\text{in/out}(\tau)$ at the input and the output of the \ac{GRU} layer in the compandor. 
As these features can have very different value ranges per unit, we normalize the features of each unit per test sequence to have an amplitude range of $-1$ to $1$ for our clustering analysis. Hence, for clustering, each frame is characterized by its normalized feature vector $\mathbf{h}_\text{in/out}(\tau)$. 

Due to the expected discriminative nature of the features, we use the L$1$ distance $\text{d}(\cdot,\cdot)$ for the $k$-means algorithm. 
Given $Q$ sources, which can be different speakers or the same speaker at different positions, we use $k=Q+1$ cluster centers, to allow one cluster for each of the $Q$ sources in a test sequence, and one cluster for signal pauses as we assume that spatial information cannot be extracted from silent signal frames.
We assign the pause label to the cluster that has the lowest signal energy in the corresponding target signal segments. The labels for the $Q$ sources are assigned by a majority vote of the assigned frames.

We quantify the clustering results in terms of the \textit{grouping success}, the average distance $\bar{d}$ of the features to their assigned cluster center, and the overall percentage of frames assigned to the pause cluster. The grouping success measures in percent how many of the frames belonging to one source are assigned to the same cluster. For $Q=2$, a grouping success of 50\% means that no distinction can be made between the two sources, while a grouping success of 100\% means that for both sources all frames belonging to the active segments of the corresponding source are assigned to the same cluster. For the calculation of the grouping success we only consider frames assigned to one of the active source clusters and disregard all frames assigned to the pause cluster. Since pauses can also appear while a speaker is active, the assignment to the pause cluster is not necessarily wrong but uninformative. Hence, we disregard these frames. 

To account for the pause cluster, we measure the overall percentage of frames assigned to the pause cluster as we expect the features after the \ac{GRU} layer to be more expressive with respect to the corresponding \ac{DoA} and to benefit from the memory of the \ac{GRU} layer. Hence, we expect that the features $\mathbf{h}_\text{out}(\tau)$ at the output of the \ac{GRU} layer can be clustered more consistently than the features $\mathbf{h}_\text{in}(\tau)$ at the input, and therefore the pause clusters should exhibit a lower percentage.

Finally, the average distance of the features to the cluster centers $\bar{d}$ is computed as 
\begin{equation}
    \bar{d} = \dfrac{1}{S}\sum_{s=1}^{S}\left(\dfrac{1}{k}\sum_{n=1}^{k}\left(\dfrac{1}{T_{n,s}}\sum_{\tau_{n,s}=1}^{T_n}\text{d}(\mathbf{c}_{n,s}, \mathbf{h}(\tau_{n,s}))\right)\right),
\end{equation}
where $\tau_{n,s} = 1,...,T_{n,s}$ indexes the frames assigned to cluster $n$ in sequence $s, s = 1,...,S$, and $\text{d}(\mathbf{c}_{n,s}, \mathbf{h}(\tau_{n,s}))$ describes the L$1$ distance from feature vector $\mathbf{h}(\tau_{n,s})$ to cluster center $\mathbf{c}_{n,s}$.
$\bar{d}$ describes how dense the clusters are: a smaller value means that the assigned features are closer to their cluster center and hence more similar to each other.

\section{Experimental Validation}
\label{sec:experiments}
So far, experiments with \ac{COSPA} have only been conducted for scenarios with one spatially static desired source \cite{Halimeh2022}. To investigate the exploitation of the spatial information contained in the multichannel signal, we perform experiments with changing speaker positions within one signal sequence. 
The purpose of our experiments is to verify the hypothesis that a layer with memory that can access all channels individually in a neural spatial filter, i.e., the \ac{GRU} layer in \ac{COSPA}'s compandor, explicitly represents spatial information about the acoustic scene at its output. 
We first examine \ac{COSPA}'s masking behavior for changing speaker positions in a scenario without background noise and with only one active speaker at any time. We analyze the features before and after the \ac{GRU} layer and measure their discriminative power via clustering. Then, we investigate whether the features distinguish between speakers or between \acp{DoA}, and, finally, we assess how robust the spatial processing is against spatially and spectrally white background noise.

\subsection{Experimental setup}
\label{subsec:setup}
For all experiments, signals are generated by simulating a room with dimensions sampled uniformly from \mbox{[4-8, 4-8, 1-4]\,m} and a reverberation time between 200\,ms and 500\,ms. A uniformly spaced linear microphone array (ULA) with $M=3$ microphones and element spacing of 0.04\,m, is randomly placed in the room with uniformly distributed positions. Defining the array's endfire directions as 0\degree\ and 180\degree, speaker positions are chosen between 0\degree\ and 180\degree.
The speech data is taken from the TIMIT database \cite{timit} with its default split between training data and test data, and convolved with the corresponding room impulse response generated according to \cite{Habets2010} for the respective speaker positions. All training and test sequences have a length of 7\,s at 16\,kHz sampling rate.

We create two training datasets of approximately 8\,h each for our experiments.
In the first dataset, denoted by \textit{DS-clean}, 
each training sequence contains two speakers which are placed randomly at $Q=2$ different positions in the room and are talking alternatingly. In each sequence, there are three segments, where Speaker~$1$ is active in the first and third segment and Speaker~$2$ is active in the middle segment. The time instants of the two speaker changes are chosen randomly, uniformly distributed between $1$ and $3$\,s, and $5$ and $6$\,s respectively. The angular distance of the two speakers' \acp{DoA} is at least $20$\degree\ and their distance to the array and the walls is at least 0.3\,m. For creating the target signal, the multichannel input speech signal image corresponding to each speaker separately is filtered with a \ac{DSB} steered towards the position of the respective speaker. The two signals for the respective speakers are combined to obtain the target signal for the end-to-end training of \ac{COSPA}. Note that this dataset does not contain any background noise and only serves the purpose of investigating the phase alignment capabilities of \ac{COSPA}.

The second training dataset, denoted by \textit{DS-WGN}, has the same setup as DS-clean but the signals consist of the speech signals and additive spatially and spectrally white Gaussian noise at \acp{SNR} between -10\,dB and 50\,dB in steps of 5\,dB.
We train \ac{COSPA} individually on both training datasets and denote the resulting trained networks as \ac{COSPA}\textsubscript{clean} and \ac{COSPA}\textsubscript{WGN} respectively.

For testing, we define four datasets. \textit{DST-clean} has the same setup as the training dataset DS-clean and consists of 50 test sequences used for the verification of the hypothesis that the GRU layer output $\widetilde{\mathbf{h}}_\text{out}(\tau)$ represents spatial information about the source. \textit{DST-WGN} has 50 test sequences per \mbox{\ac{SNR} $\in [-10, 5, 0, 5, 10, 20, 30, 50]$\,dB} but otherwise has the same setup as the training dataset DS-WGN. This test dataset is used to test the robustness against noise of the spatial representation in the \ac{GRU} layer.
To show that the features $\widetilde{\mathbf{h}}_\text{out}(\tau)$ encode spatial and not speaker information, we use \textit{DST-1Pos}, which contains 50 sequences with two alternating speakers placed at the same position in the room, and \textit{DST-1Spk}, which contains 50 signals from a single speaker including one position change. $Q=2$ and $k=3$ for all experiments. 

We parameterize \ac{COSPA} as published in \cite{Halimeh2022}. Especially, the fully connected layer before the \ac{GRU} layer in the compandor and the \ac{GRU} layer itself both have $U_\text{in} = U_\text{out} = 128$ units generating both the real and imaginary part of the $U_\text{in/out}$ complex-valued features $\widetilde{h}_\text{in/out}(\tau,u)$. For computing  the \ac{STFT} features as input to \ac{COSPA}, we use frames of 1024 time-domain signal samples and a frame shift of 512 samples.

Since clustering is initialization-dependent, we report the results for the clustering as average over five trials, where in each trial the best out of five $k$-means clustering attempts is used to compute the evaluation metrics.
\subsection{Results}
\label{subsec:results}
\begin{figure}[t]
\centering
        \includegraphics[height=0.37\linewidth, width=\linewidth]{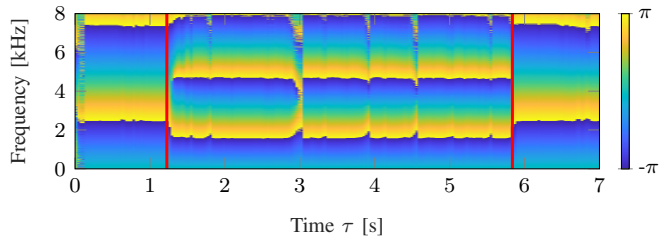}
    
\caption{Exemplary phase component of the estimated mask $\mathcal{M}_2$. Speaker~$1$ is at $103$\degree\ and Speaker~$2$ at $43$\degree. The red lines indicate the change of the active speaker.}
\label{fig:phase}

\end{figure}%
\begin{figure}
    \input{sample_18_in_features.tikz}
    
    \vspace{-.1cm}
    \input{sample_18_out_features.tikz}
        	
    \vspace{-.1cm}
    \input{sample_18_in_clusters.tikz}
        	
    \vspace{-.1cm}
    \input{sample_18_out_clusters.tikz}  	
    \vspace{-.1cm}%
%
%
\begin{subfigure}[t]{\columnwidth}
\subcaption{Target signal}
\begin{tikzpicture}
	\begin{axis}[%
	width=7cm,
	height=1.5cm,
	 ylabel near ticks,
	 xlabel near ticks,
	scale only axis,
	point meta min=-1,
	point meta max=1,
	axis on top,
	xmin=0,
	xmax=7,
	ymin=-1,
	ymax=1,
	ylabel style={font=\color{white!15!black}},
	ylabel={\footnotesize Target signal},
	ticklabel style={font=\footnotesize},
    ytick={-1, 1},
    yticklabels={\scalebox{0.75}[1.0]{$-$}$1$, $1$},
    xlabel={\footnotesize Time $\tau$ [s]},
	]
	\addplot [forget plot] graphics[xmin=0, xmax=7, ymin=-1, ymax=1] {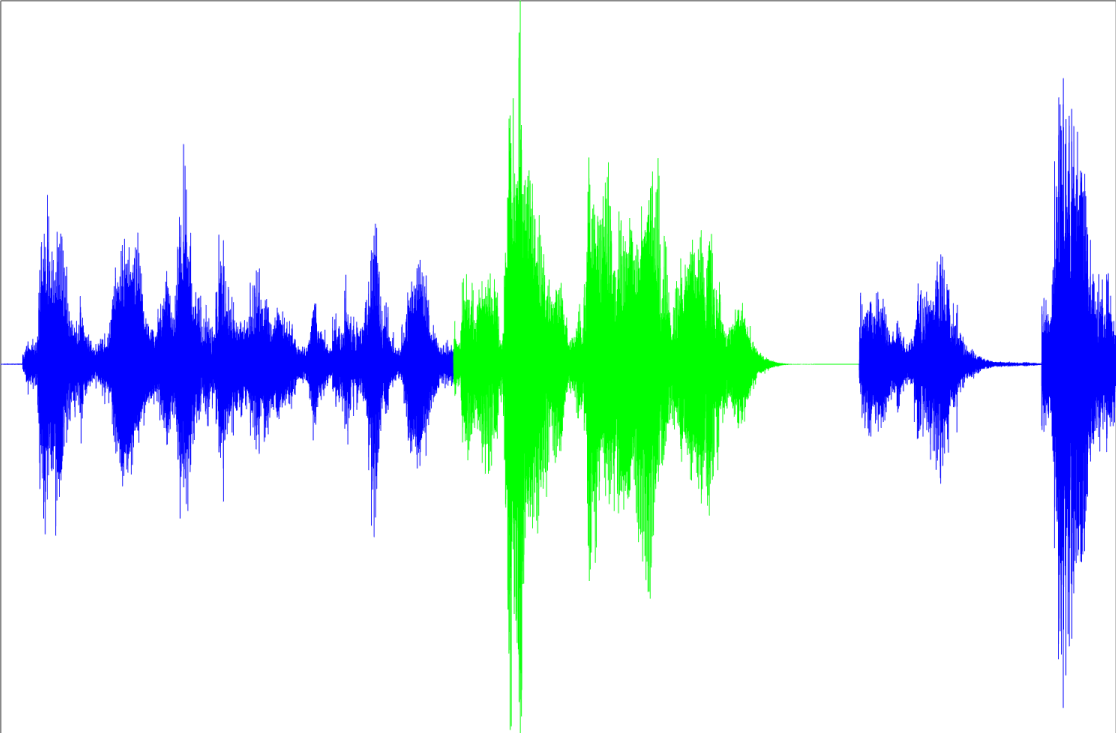}; 
	\addplot [color=red, line width=0.4mm, forget plot]
	  table[row sep=crcr]{%
	2.8421	-1\\
	2.8421 1\\
	};
	\addplot [color=red, line width=0.4mm, forget plot]
	  table[row sep=crcr]{%
	5.3825	-1\\
	5.3825	1\\
	};
	\end{axis}
\end{tikzpicture}%
\label{subfig:target}
\end{subfigure}
        	
    \caption{Stacked real and imaginary part of exemplary feature vectors $\mathbf{h}(\tau) $ in COSPA\textsubscript{clean} before (a) and after (b) the GRU layer, and the respective clustering result (P: Pause, $1$: Position $1$, $2$: Position $2$) (c, d) and time-domain target signal (e). Speaker~$1$ is at $51$\degree, Speaker~$2$ at $78$\degree. The red lines indicate the position change.}
    \label{fig:clusters} 	
\end{figure}%
\paragraph*{Spatial adjustment}
We first investigate whether \ac{COSPA} can react to changing spatial conditions within an acoustic scene without guiding information during testing. As a typical example, in Fig.~\ref{fig:phase}, it can be seen that for the test dataset DST-clean the phase mask $\mathcal{M}_2$ generated by \ac{COSPA}\textsubscript{clean} indeed reflects the \ac{DoA} change following the changing speaker activity. From this it can be concluded that spatial information of the received signal is used by \ac{COSPA} to generate the phase component of the estimated mask $\mathcal{M}_m$.

Based on this observation and the special role of the \ac{GRU} layer in \ac{COSPA}'s compandor (cf.~Sec.~\ref{sec:method}), we investigate whether the features generated before and after the \ac{GRU} layer also reflect the positional changes in the signal.

\paragraph*{Features}
Figs.~\ref{fig:clusters}\subref*{subfig:features_input} and \subref*{subfig:features_output} show the feature vectors $\mathbf{h}_\text{in}(\tau)$ and $\mathbf{h}_\text{out}(\tau)$ at the input and output of the GRU layer, respectively, corresponding to the time-domain signal in Fig.~\ref{fig:clusters}\subref*{subfig:target}. In Fig.~\ref{subfig:features_input}, speech pauses are already clearly discernible from speech for $\mathbf{h}_\text{in}(\tau)$, especially around $\tau=5s$ and following $\tau=6s$. Otherwise, no distinctive differences between the two speakers are apparent. In Fig.~\ref{subfig:features_output}, it can be observed that the features $\mathbf{h}_\text{out}(\tau)$ are more consistent along the time axis than $\mathbf{h}_\text{in}(\tau)$, that pauses are still discernible, and that there is clearly a specific feature pattern for each speaker.

\paragraph*{Clustering results}
Figs.~\ref{fig:clusters}\subref*{subfig:cluster_input},\subref*{subfig:cluster_output} show the clustering results according to the features presented in Figs.~\ref{fig:clusters}\subref*{subfig:features_input},\subref*{subfig:features_output}. Clusters are colored according to the signal part that is represented. It can be clearly seen that for the input features $\mathbf{h}_{\text{in}}(\tau)$ clustering appears to be uninformative concerning the speaker position, while the pauses are detected quite well. For the output feature vectors $\mathbf{h}_{\text{out}}(\tau)$ clustering is very informative regarding the activity of a source from a certain DoA. 

\begin{table}[t]
\setlength{\tabcolsep}{5pt} 
\renewcommand*{\arraystretch}{1.1}
\centering
\caption{Grouping success and characteristics of clusters for the \ac{COSPA}\textsubscript{WGN} features before and after the \ac{GRU} layer.} 
\scalebox{0.86}{
\begin{tabular}{lc|>{\centering}m{0.7cm}>{\centering}m{0.7cm}|>{\centering}m{0.9cm}>{\centering}m{0.9cm}|>{\centering}m{0.9cm}c}
\toprule
Test & SNR & \multicolumn{2}{c|}{grouping success} & \multicolumn{2}{c|}{average distance to} & \multicolumn{2}{c}{relative size of} \\ 
dataset &{[dB]} & \multicolumn{2}{c|}{[\%] $\uparrow$} &  \multicolumn{2}{c|}{cluster centers $\bar{d}$ $\downarrow$} & \multicolumn{2}{c}{pause cluster [\%] $\downarrow$} \\
\hline
& & $\mathbf{h}_\text{in}$ & $\mathbf{h}_\text{out}$ & $\mathbf{h}_\text{in}$ & $\mathbf{h}_\text{out}$ & $\mathbf{h}_\text{in}$ & $\mathbf{h}_\text{out}$ \\
\hline
DST-clean & $\infty$ & 78.3 & 95.1 & 37.5 & 33.3 & 12.5 & 8.9\\
DST-1Spk & $\infty$  & 71.0 & 94.6 & 39.4 & 34.6& 28.8 & 19.1\\
DST-1Pos & $\infty$  & 57.7 & 66.3 & 37.4 & 32.6 & 30.2 & 24.9 \\
\hline
\multirow{8}{*}{DST-WGN} &50 & 80.7& 94.5 & 37.6 & 33.2 & 32.0 & 21.5 \\
&30 & 80.4 & 95.3& 38.9 & 34.2 & 31.7& 21.9 \\
&20 & 79.1 & 94.6 & 41.2 & 35.4 & 33.0 & 24.8 \\
&10 & 74.3 & 93.3 & 41.6 & 34.6 & 41.7& 27.3 \\
&5 & 68.9 & 90.1&42.0 &33.7 & 45.2 & 31.4\\
&0 & 66.0& 84.4 & 42.6 & 32.6 & 48.6 & 36.6\\
&-5 & 63.5 & 82.0 & 43.6 & 31.8 & 45.0 & 38.9\\
&-10 & 61.1 & 77.4 & 44.9 & 32.5 & 32.6 & 29.4 \\
\bottomrule
\end{tabular}
}
\label{Table:Clustering}
\end{table}

Table~\ref{Table:Clustering} provides the clustering results for features obtained for \ac{COSPA}\textsubscript{WGN} for the four test datasets. For DST-clean, it can be seen that 95.1\% of frames belonging to either speaker position are assigned to the corresponding cluster at the output of the \ac{GRU} layer, whereas only 78.3\% of corresponding frames are grouped before the \ac{GRU} layer. This shows that the memory in the \ac{GRU} layer supports extracting the spatial information significantly, but also that the features after the first fully-connected layer in the compandor already contain a limited amount of spatial information. 

Table~\ref{Table:Clustering} also gives the average distance of the data points to the respective cluster centers $\bar{d}$, which confirms that the clusters for $\mathbf{h}_\text{out}(\tau)$ are more compact than those for $\mathbf{h}_\text{in}(\tau)$, since $\bar{d}$ is smaller. This also points to more informative and discriminative features after the \ac{GRU} layer.

\paragraph*{The pause cluster}
It can be noted from both Fig.~\ref{fig:clusters} and Table~\ref{Table:Clustering} that fewer frames are assigned to the pause cluster at the output of the \ac{GRU} layer than at the input, and that those frames belong to longer speech pauses and are not just single frames. This can be explained by the fact that at the input of the \ac{GRU} layer, the pause cluster aside from the silent frames also contains several undecided frames and therefore has relatively many frames assigned to it. At the output of the \ac{GRU} layer this cluster contains only frames with actual speech pauses and hence also has a lower number of frames assigned to it. This also shows the beneficial effect of the \ac{GRU} layer's memory, since it creates, based on previous signal parts, more meaningful features for frames where signal energy is low  (cf.~Figs.~\ref{fig:clusters}\subref*{subfig:features_input},\subref*{subfig:features_output}). Before the \ac{GRU} layer, these frames would be assigned to the pause cluster or, even worse, to the wrong speaker cluster.
Note that since the grouping success is only computed based on frames sorted into one of the speaker clusters, erroneously assigning many low-level speech frames to the pause cluster can lead to relatively high grouping success scores even at the input of the \ac{GRU} layer. 

\paragraph*{Adaptation to \ac{DoA} changes}
We observed in our experiments that the grouping success for the frames directly following a position change depends on whether the position change coincides with a speech pause or not. As an example, in Fig.~\ref{subfig:cluster_output} the second switch around $\tau=5.4$\,s coincides with a switch from a pause to a speech signal. \ac{COSPA} is able to identify this change immediately. If the change had happened at $\tau=4.9\,$s, \ac{COSPA} would not know that the position of the desired speaker has changed, because this will only become apparent when the speaker starts talking. Unsurprisingly and similar to acoustic source localization, adaptation to a new DoA cannot be expected during pauses.
A similar effect can be observed for a switch between two \acp{DoA} during speech activity. In some scenarios the assignment of the input features after this \ac{DoA} switch changes almost instantaneously, while for the output features changing the cluster assignment takes a few frames when speech is present. This is expected behavior, since the memory provided by the GRU will remember the spatial setup of the scene and hence will wait for more evidence for a \ac{DoA} change before reassigning the features. The delay in changing the features is beneficial for increasing robustness against directional interferers, e.g., when applying \ac{COSPA} to target speaker extraction and speech enhancement in the presence of multiple sources. (Note that this behavior depends also on the differences in amplitude of the signals and cannot be observed in Fig.~\ref{fig:clusters}\subref*{subfig:cluster_output} at $\tau=2.9$\,s as the signal amplitudes differ notably.)

\paragraph*{Source independence of features}
To validate that the features generated by the \ac{GRU} layer indeed reflect different spatial setups and not different speakers as they are present in DST-clean, we also test \ac{COSPA}\textsubscript{WGN} on DST-1Spk and DST-1Pos, which contain signals from a single speaker with changing position or from two speakers at the same position respectively (cf.~Sec.~\ref{subsec:setup}). As can be seen in Table~\ref{Table:Clustering}, for DST-1Spk the clustering scores for $\mathbf{h}_\text{out}$ are good even though only one speaker is present. On the other hand, the grouping success for $\mathbf{h}_\text{out}$ for DST-1Pos is really low, which shows that the two speakers at the same position cannot be distinguished by the features of the \ac{GRU} layer. These results show that \ac{COSPA} indeed reacts to a change of spatial setup and not to the characteristics of a different speaker.

Apart from the results shown here, we find that \ac{COSPA} can also adapt to a different number of position changes during testing than it was trained on. This underscores the flexibility of such a neural spatiospectral filter.

\paragraph*{Robustness against noise}
So far, the results have been presented for \ac{COSPA} tested on signals without any background noise. To demonstrate how well the spatial information is captured in the \ac{GRU} layer in the presence of noise and for the task of noise reduction rather than just phase alignment, we provide the clustering test results for \ac{COSPA}\textsubscript{WGN}, which was trained for a wide range of \ac{SNR} levels, for test sequences with various \ac{SNR} levels in Table~\ref{Table:Clustering}. It can be seen that for all \acp{SNR} both the grouping success and the cluster density are notably better for the output than for the input features of the \ac{GRU} layer. As expected, the grouping success decreases for lower \acp{SNR}. The number of frames assigned to the pause cluster also decreases from input to output of the \ac{GRU} layer, which points to more discriminative features and a good feature memory in the \ac{GRU} layer. 

In summary it can be stated that the spatial information exploited for spatiospectral filtering with \ac{COSPA} is processed by the \ac{GRU} layer in the compandor and clearly expressed in the output features of this \ac{GRU} layer. From the exemplary character of \ac{COSPA}'s single \ac{GRU} layer in an otherwise memoryless neural network, we can infer that the \ac{GRU} layer indeed implements a noise-robust and source signal-independent phase equalization for coherent multichannel signals.  

\section{Conclusion}
\label{sec:conclusion}
In this paper, we investigated, through the example of \ac{COSPA}, the capabilities of a neural spatiospectral filter to represent and exploit spatial information. We found that the complex-valued \ac{GRU} layer in \ac{COSPA}'s compandor processes the spatial information and that the features obtained at the output of the \ac{GRU} layer provide discriminative information about the \ac{DoA} of a signal. Different from a \ac{DSB} and as long as only one point-like speech source is active, \ac{COSPA} does not require information about the \ac{DoA} of the signal to adapt to positional changes of the source. Furthermore, we showed that clustering of input and output features is an appropriate tool for assessing the capabilities of a neural network layer for the task of spatial filtering.

\bibliographystyle{IEEEbib}
\bibliography{references}

\end{document}